\documentclass[sigconf]{acmart}

\usepackage{booktabs}

\usepackage{balance}

\def\BibTeX{{\rm B\kern-.05em{\sc i\kern-.025em b}\kern-.08emT\kern-.1667em\lower.7ex\hbox{E}\kern-.125emX}}
    
\setcopyright{acmcopyright}

\begin{document}

\copyrightyear{2019} 
\acmYear{2019} 
\acmConference[MADiMa '19]{5th International Workshop on Multimedia Assisted Dietary Management}{October 21, 2019}{Nice, France}
\acmPrice{15.00}
\acmDOI{10.1145/3347448.3357169}
\acmISBN{978-1-4503-6916-9/19/10}
%
\title{Flavour Enhanced Food Recommendation}

%
\author{Nitish Nag}
\email{nagn@uci.edu}
\affiliation{%
  \institution{University of California, Irvine}
  \city{Irvine}
  \state{California}
  \country{United States of America}
  \postcode{92697}
}

\author{Aditya Narendra Rao}
\email{13aditya13@gmail.com}
\affiliation{%
  \institution{PES University}
  \streetaddress{100 Feet Ring Road}
  \city{Bangalore}
  \state{Karnataka}
  \country{India}
  \postcode{560085}
}

\author{Akash Kulhalli}
\email{akash.kulhalli@gmail.com}
\affiliation{%
  \institution{PES University}
  \streetaddress{100 Feet Ring Road}
  \city{Bangalore}
  \state{Karnataka}
  \country{India}
  \postcode{560085}
}
  
\author{Kushal Samir Mehta}
\email{kushalmehta13@gmail.com}
\affiliation{%
  \institution{PES University}
  \streetaddress{100 Feet Ring Road}
  \city{Bangalore}
  \state{Karnataka}
  \country{India}
  \postcode{560085}
}
 
\author{Nishant Bhattacharya}
\email{nishantb21@gmail.com}
\affiliation{%
  \institution{PES University}
  \streetaddress{100 Feet Ring Road}
  \city{Bangalore}
  \state{Karnataka}
  \country{India}
  \postcode{560085}
}

\author{Pratul Ramkumar}
\email{pratul.ramkumar@gmail.com}
\affiliation{%
  \institution{PES University}
  \streetaddress{100 Feet Ring Road}
  \city{Bangalore}
  \state{Karnataka}
  \country{India}
  \postcode{560085}
}

\author{Aditya Bharadwaj}
\email{aditya21nov@gmail.com}
\affiliation{%
  \institution{PES University}
  \streetaddress{100 Feet Ring Road}
  \city{Bangalore}
  \state{Karnataka}
  \country{India}
  \postcode{560085}
}

\author{Dinkar Sitaram}
\email{dinkars@pes.edu}
\affiliation{%
  \institution{PES University}
  \streetaddress{100 Feet Ring Road}
  \city{Bangalore}
  \state{Karnataka}
  \country{India}
  \postcode{560085}
}

\author{Ramesh Jain}
\email{jain@ics.uci.edu}
\affiliation{%
  \institution{University of California, Irvine}
  \city{Irvine}
  \state{California}
  \country{United States of America}
  \postcode{92697}
}

%
\renewcommand{\shortauthors}{Aditya et al.}

%
\begin{abstract}
We propose a mechanism to use the features of flavour to enhance the quality of food recommendations. An empirical method to determine the flavour of food is incorporated into a recommendation engine based on major gustatory nerves. Such a system has advantages of suggesting food items that the user is more likely to enjoy based upon matching with their flavour profile through use of the taste biological domain knowledge. This preliminary intends to spark more robust mechanisms by which flavour of food is taken into consideration as a major feature set into food recommendation systems. Our long term vision is to integrate this with health factors to recommend healthy and tasty food to users to enhance quality of life.
\end{abstract}

%
%



%
\maketitle

\begin{figure}
\small
\centering
\includegraphics[width=1 \columnwidth]{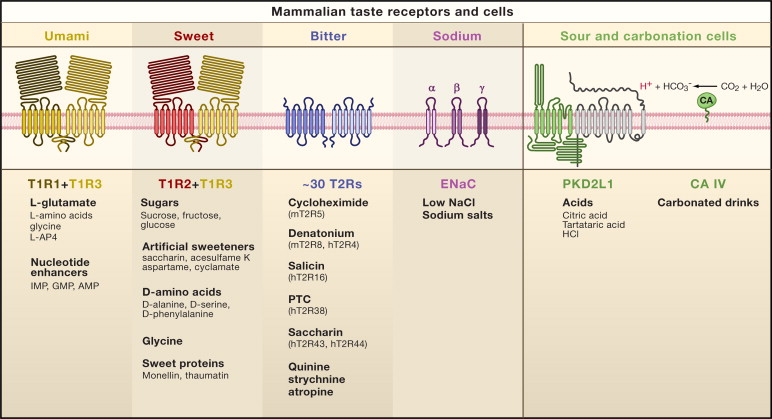}
\caption{Mammalian taste receptors are the basic unit of biological taste. Understanding the features which impact these receptors is the fundamental start to profiling users and food in their culinary preferences. As biological discoveries in protein structure and function of taste are uncovered, the knowledge must be incorporated into the recommendation systems and health navigation paradigms \cite{Yarmolinsky2009}.}~\label{fig:receptors}
\vspace{-5mm}
\end{figure}

\begin{figure*}
\centering
\includegraphics[width=3.5in]{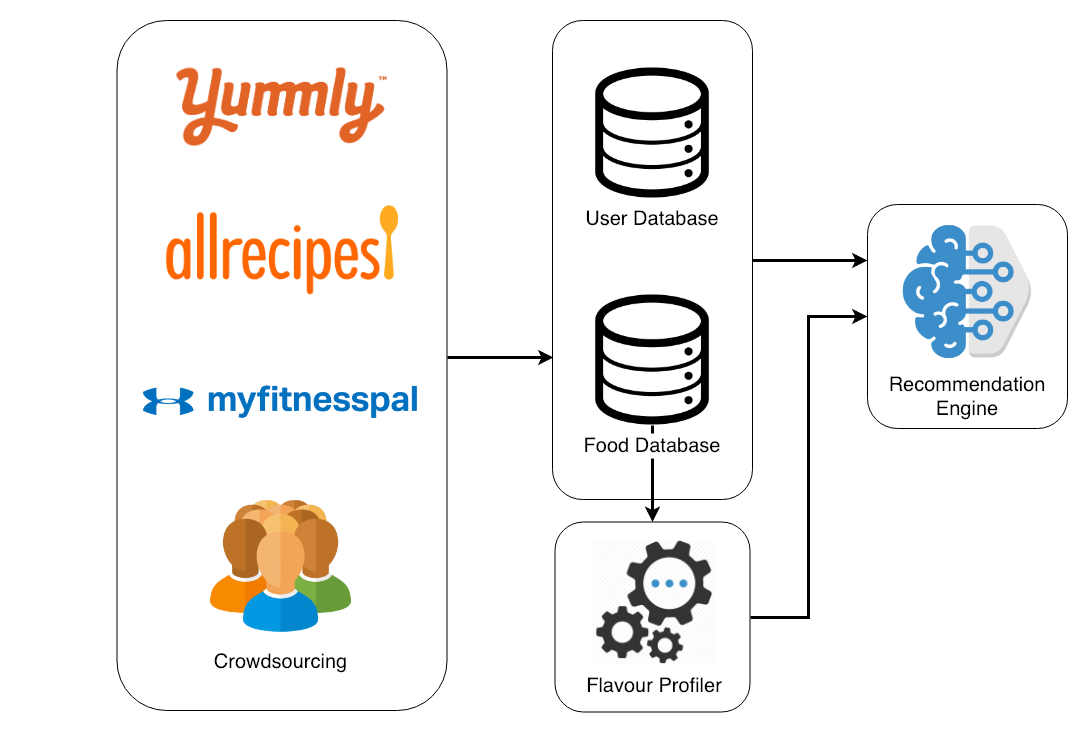}
\caption{Platform Architecture}
\label{fig:platformarch}
\end{figure*}

\section{Introduction}
Food is essential for human life. Beyond sustaining health by providing nourishment, it is also an integral part of human culture and quality of life. Modern information about nutrition and food fosters development of food computing methods \cite{min2018survey}. The notion of food computing involves obtaining food data and identifying areas where it can be applied effectively, such as health, food sciences, culinary art, and human behaviour. Food computing collects data from multiple sources and involves tasks such as perception, recognition, retrieval, recommendation, prediction and monitoring of food intake. One of the key outcomes of food computing is understanding the relationship between dietary choices and health state \cite{nag2019navigational, nag2018cross}. A healthy diet promotes overall well-being and lowers the risk of chronic diseases. To aid in building a healthy diet, algorithms can potentially compute health scores for food items based on the users health status \cite{nag2019navigational, nag2018cross}, the item nutritional features, along with context and other environmental factors \cite{nag2017live}. However, healthy food suffers from the adoption problem, since healthy food can be in conflict with taste preferences \cite{forwood2013choosing}. This perception results in people consuming food without full regard for the effects on their overall health. In this work, we describe a preliminary recommendation system that incorporates taste preferences. We determine to what extent the characteristics of dishes, namely flavour and cuisine, and user inclination affects the quality of food recommendations. In combination, we hope this concept is used to help individuals make health-aware dietary choices that are enjoyable.

\section{Related Work}

Modern recommendation systems proactively identify and provide a user with ranked search results that are context-aware. Earlier food recommendation engines use TF-IDF to generate vectors from food items while taking into account food database information \cite{el2012food}. An input to these systems is generally a question similar to "what\textquotesingle s for lunch". Evaluation of these early systems used traditional metrics such as accuracy, precision, and recall. Unfortunately these metrics fail to capture the quality of recommendations in relationship to real world implementation for enjoyment or health. Effectively extending the food recommendation to incorporate the individual health state criteria and culinary flavour and user preferences will be the next evolution of more personalized food recommendation \cite{freyne2010intelligent, freyne2011recipe, elsweiler2015bringing, nag2019navigational, nag2018cross}. Qualitative analysis based on user feedback will also be essential to improving quality of recommendations. Approaches where users are clustered into distinct groups unfortunately may reduce the personal level recommendation \cite{svensson2000recipe}. Other approaches involve various levels of personalizing outputs for a user with a unique focus. For instance some evaluations are more focused on usability and appeal of the recommendations, user previously specified interests, or based on a certain demographics \cite{elahi2015interaction, kuo2012intelligent, aberg2006dealing}. Furthermore, perceptions of what is healthy or tasty vary greatly among people \cite{Ofli2017}. Many of these research efforts have primarily focused on meeting dietary and nutritional constraints. This is why we see value in building the research body on flavour characteristics.

\begin{table}
  \caption{Database Statistics}
  \label{tab:stat}
  \begin{tabular}{cc}
    \midrule
    User Database, Total reviews& 30,193 \\
    User Database, Unique Users& 22,625 \\
    User Database, Users with greater than 5 reviews& 466\\
    Food Database, Total Dishes &1381 \\
    Food Database, Indian Dishes& 1051 \\
  \bottomrule
\end{tabular}
\end{table}

\section{Overview}

Like most other recommendation systems, ours uses two
database components of items and users to form the input. Food items were curated containing their ingredient list, nutritional values and their cuisines were obtained from public food websites. We target food items that reflect the diet of the South Asian audience. Additionally, to account for the regional variety, food items were crowd-sourced by sending out surveys to 200 users.

This resulted in a food database containing 1381 items. To account for missing values, we used a nutrition facts database (MyFitnessPal) to fill in the gaps for nutritional values, while the cuisines were manually tagged. The second component of user profiles was constructed by obtaining user reviews for the food items that were previously scraped from public domain websites and crowd-sourced ratings for the most recent items consumed. User reviews were then used to build a user preferences profile. For each food item in the curated database, we estimate the intensity of the five basic flavours of sweet, salty, richness, umami and bitter on a scale of 1 to 10, 10 indicating highest intensity. This is done by identifying and quantifying the most influential chemicals for each flavour from the dish. We did not address the basic taste of sour in this work due to challenges in empirically profiling the flavour from the data we gathered.

Once the flavour scores for all food items have been generated, we consider it as an additional feature for the recommendation engine. This means that each food item will have the five flavour scores as five extra dimensions. We then apply a similarity score to predict how much a user will rate another food item based on previous ratings.

\section{Implementation}
\subsection{\textbf{Flavour Computing}} \label{AA}
An objective flavour metric is a scale that looks purely at the content of the food item without considering any external factors such as user/cooking preferences. A prime example of this is the Scoville scale, which measures the heat of spiciness of chilli peppers based on the amount of capsaicin present in it. However, no such metric exists for other flavours. Attempts have been made to establish a relationship between nutrients and taste \cite{van2012taste}. We follow a similar approach and attempt to identify the elements that influence each flavour the most based on biological receptor functioning as shown in Figure \ref{fig:receptors}. In all the calculations mentioned below, the total weight considered is the Total Active Nutrient Weight (in grams) (TANW) of the dish.

\begin{enumerate}
    \item Salt:
    
    The quantity of sodium indicates the saltiness of a dish. To highlight the prominence of sodium in the dish relative to its weight, we identify the ratio between the total sodium present in the dish and the total nutritional weight. This value is then normalized for soduium chloride, by using the molar masses of 22.99 and 35.45 g/mol respectively, 100 g of NaCl contains 39.34 g Na and 60.66 g Cl.
    \begin{equation}
        \label{salteqn}
        Salt\ Score\ =\ \frac{100}{38.758} \cdot \frac{Sodium\ content (g)}{TANW}
    \end{equation}
    
    \item Sweet:
    
    The carbohydrate content in foods consists of monosaccharides, disaccharides and polysaccharides. Monosaccharides (glucose, fructose) and disaccharides (sucrose) contribute positively towards the sweetness of a dish, while indeigestable polysaccharides (dietary fibres) negate their effects. We do a weighted addition of the two components to compute the sweetness score.
    \begin{gather*}
        Sweet(A)=\frac{sugar (g)-fibre (g)}{TANW} \\
        Sweet(B)=\frac{sugar (g)}{carbohydrates (g)}
    \end{gather*}
    \begin{equation}
        \label{sweeteqn}
        Final\ sweet\ score=x \cdot A+y\cdot B
    \end{equation}
    where $x$ and $y$ are 0.85 and 0.1, respectively. \\
    
    \item Bitter:
    
    Bitterness is indicated by the calcium and iron content in the dish. Although this is a crude metric since bitterness is quite complex, we feel this is an adequate start to profiling bitterness given data availability. To combat the lack of data available for these in Indian dishes, we maintained a list of ingredients that were manually tagged as \textquotesingle bitter\textquotesingle or \textquotesingle too bitter\textquotesingle. The rank of each ingredient for each group is added and a weighted addition is performed, taking into account the iron content.
    \newline
    \begin{equation}
        \label{bittereqn}
        \begin{split}
            Bitter\ score=x \cdot \sum_{0}^{k}(bitter\ value)\ + \\
            y\cdot\sum_{0}^{k'}(too\ bitter\ value)+z \cdot iron\ content (g)
        \end{split}
    \end{equation}
    where $x$, $y$, and $z$ are weights with the values 0.8, 2.4 and 1.3, while $k$ and $k'$ are the number of ingredients tagged as bitter and too bitter respectively. The values for $bitter$ are assigned to ingredients based on a relative intensity of bitterness, normalized to 1.\\
    \item Umami:
    
    The umami taste is biologically sensed by the amino acid glutamate receptor, which is prominent in protein-rich dishes. The ingredients were divided into groups like meats, vegetables, umami enhancers (MSG), and protein supplements (whey), sorted in order of their glutamate and protein content. A multiplier was assigned to each ingredient group, which was then totalled to obtain the final group score. The fractional protein content and group score were again subjected to a weighted addition, obtaining the final umami score.
    \begin{gather*}
        Umami(A)=\frac{Protein (g)}{TANW} \\
        Umami(B)=\sum_{}{}\{group\ multiplier\ *\ group\ score\}
    \end{gather*}
    \begin{equation}
        \label{umamieqn}
        Umami\ score=A+B
    \end{equation}
    
    where $group\ multiplier$ is the weight of each of the following groups - Protein supplements, vegetables, meat and savoury phrases, with their respective multipliers being 0.8, 7, 3 and 9.45. \\
    
    \item Richness:
    
    The richness score is computed by considering saturated fats, cholesterol and total fats. The saturated fat content is used as a fraction of the total fat content present in the food item. The ratio of the total fat content to the total active nutritional weight and the amount of cholesterol in the dish relative to its weight is also taken into account while calculating the richness score. The final score is a linear combination of the aforementioned factors. Although richness is not considered a standard taste receptor at the moment, culinary arts have long used it as a factor in designing and cooking dishes.

\begin{displaymath}
            A=\frac{saturated \ fat (g)}{fat (g)}
        \end{displaymath}
        \begin{displaymath}
            B=\frac{fat (g)}{TANW}
        \end{displaymath}
        \\
        \begin{displaymath}
            C=\frac{cholesterol (mg)}{TANW} \ \cdot \ 1000
        \end{displaymath}
        \\
        \begin{displaymath}
            D=(\ x \ \cdot \ A ) \ + \ (y \ \cdot \ B) \ + \ (z \ \cdot \ C)
        \end{displaymath}
        \\
        \begin{equation}
            \label{richeqn}
            Richness\ score=\frac{D}{0.992}\cdot10
        \end{equation}
        \\
The weights $x, y$ and $z$ are 0.5, 0.7, 50 respectively, and were arrived at via matching with user descriptions of richness.
\end{enumerate}

\begin{table}
  \caption{Score Samples}
  \label{tab:score_samp}
  \begin{tabular}{cccccc}
    \toprule
    Dish & Bitter & Rich & Salt & Sweet & Umami\\
    \midrule
    Curried bean salad & 0.961 & 0.7 & 2.63 & 2.47 & 2.534\\
    Aloo phujia& 2.149& 2.3 & 3.116 & 0.27 & 9.271\\
    Palak paneer& 1.436& 2.25 & 1.184 & 1.12 & 8.064\\
    Channa masala& 2.012& 2.79 & 3.41 & 0.88 & 9.538\\
    Cilantro pesto & 0.604 & 4.45 & 0.904 & 0.57 & 2.198\\
    \bottomrule
\end{tabular}
\end{table}

Sample scores are shown in Table \ref{tab:score_samp}. For each of the five food items, flavour scores are indicated on a scale of 1 to 10. \\

To validate this system, we conducted a survey of 150 users. We built a website where the users were required to assign flavour scores for randomly sampled dishes. The survey entries were used as an input to the validation system, along with the flavour scores generated by the system. It then computes the error which indicates the difference between the system-generated and the user-provided scores. Note that error is 0 if abs(variance) < ACTION THRESHOLD.

\begin{figure*}[t]
    \centering
    \includegraphics[width=5in]{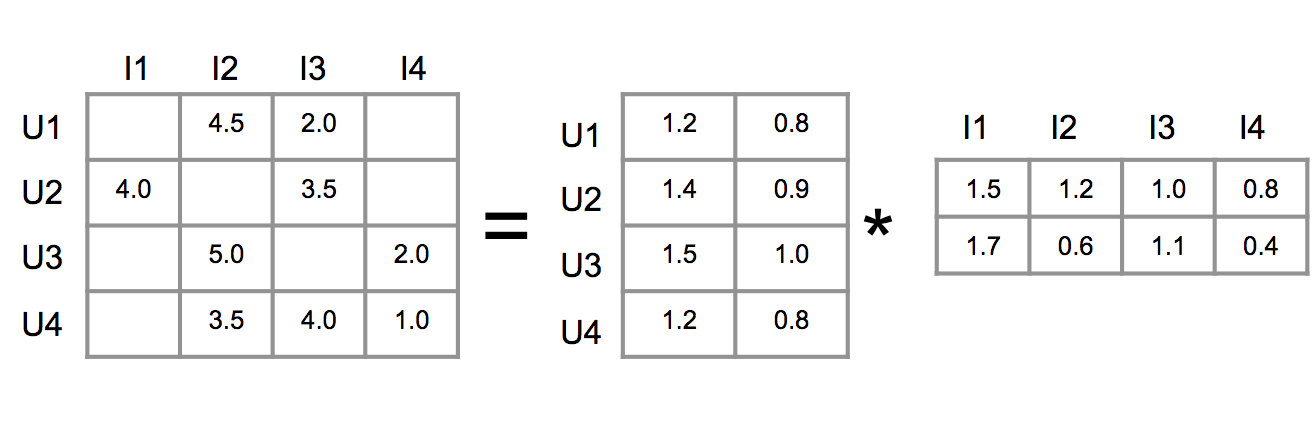}
    \caption{User-Item and latent matrices}
    \label{fig:matrices}
    \end{figure*}

\begin{displaymath}
    \label{taste_error}
    error = Q_3\cdot\ln{(variance)}
\end{displaymath}
\begin{equation}
    Final\ taste\ score=Generated\ taste\ score-error
\end{equation}

Here, $Q_3$ is the upper quartile of the list of results obtained from the survey. ACTION\ THRESHOLD is a tuned value above which the error correction is activated per taste. This is done to account for minor user-to-user variations. This process is repeated per taste to obtain scores adjusted for user feedback.

Here, the variance is computed differently than the conventional procedure - it is computed on the data list obtained by computing the difference between the generated and the surveyed taste scores. This, therefore, provides the actual variation between the generated scores and what the surveyed users expect.
Including the upper quartile ensures the majority (75\%) of user responses are accounted for, while avoiding the outliers, such as responses that may go against the general consensus. An example of such a response is a user whose taste preference is significantly skewed towards a particular flavour.

\begin{table}
  \caption{Obtained Error Values}
  \label{tab:error_values}
  \begin{tabular}{ccl}
    \toprule
    Taste & Error \\
    \midrule
    Bitter & 1.25 \\
    Rich & -0.45 \\
    Salt & -0.38 \\
    Sweet & 0.02 \\
    Umami & 8.64 \\
  \bottomrule
\end{tabular}
\end{table}

    \begin{table}
      \caption{Sample vector (Truncated)}
      \label{tab:vector_values}
      \begin{tabular}{cccccc}
        \toprule
       Dish\textbackslash Tags & Potato & Spinach & Flour & Paneer & ...\\
        \midrule
        Aloo Paratha & 0.877 & 0 & 0.685 & 0 & ... \\
        Palak Paneer & 0 & 0.819 & 0 & 0.841 & ... \\
      \bottomrule
    \end{tabular}
    \end{table}
    
    \begin{table}
      \caption{Same as Table \ref{tab:vector_values}, including taste scores (Truncated)}
      \label{tab:vector_values_with_flavour}
      \begin{tabular}{ccccccc}
       \toprule
       Dish\textbackslash Tags & Potato & Spinach & ... & Bitter & Salt & ... \\
        \midrule
        Aloo Paratha & 0.877 & 0 & ... & 2.29 & 3.83 & ... \\
        Palak Paneer & 0 & 0.819 & ... & 3.72 & 3.02 & ... \\
      \bottomrule
    \end{tabular}
    \end{table}
    
The error in Table \ref{tab:error_values} has been computed over the food data set for all users. The values as indicated in the table above shows the sweet, salty, bitter and rich flavour scores generated by our system are in line with the general consensus of the surveyed users. However, there is a significant deviation from the user-rated scores and the system-generated scores in the case of umami. A possible explanation for this anomaly could be due to the fact that a good understanding of the umami flavour is lacking among the general populace of our study. If this data is looked at on a per-user basis, it behaves as a sensitivity factor for each flavour. This way, a profile of the flavour sensitivity can be developed, that can be used to personalize the recommendation of food items even further.

\subsection{\textbf{Recommendation Engine}}
In our work we explore two types of recommendation systems - Collaborative Filtering and Content Based Filtering.
With an aim to incorporate food flavours to improve the quality of recommendations, we compare and contrast the effects of including flavour when making food recommendations.

\begin{enumerate}
    \item \textbf{Collaborative Filtering:}
    
    The Collaborative Filtering approach for recommendation looks to make predictions regarding a user\textquotesingle s preference by collecting preferences from multiple similar users. The assumption in Collaborative Filtering is that people who view and evaluate items in a similar fashion are likely to assess other food dishes in a similar manner.
    
    Matrix Factorization is a Collaborative Filtering algorithm that takes as input a User-Item Rating Matrix. This matrix is sparse since it is not likely that a user has rated all dishes in the food database. The approach aims to break down the User-Item matrix into two matrices of latent user and item representation. The intent of this approach is to reform the original User-Item matrix while filling in the missing ratings. Figure \ref{fig:matrices} depicts a User-Item Matrix and latent matrices, which when multiplied, yield predicted scores for items a user has not rated while trying to generate scores as close to the original score for items the user has rated.

    \begin{table}
    \caption{Dishes reviewed by the user}
    \label{tab:user_reviewed_dishes}
    \centering
    \begin{tabular}{cc}
    \toprule
        Dish Name & Rating \\
    \midrule
        Chole Bhature & 4 \\
        Paneer Tikka Masala & 5 \\
        Veg Biryani & 4 \\
        Bisi Bele Bath & 2 \\
        Aloo Paratha & 3 \\
        Vegetarian Korma & 2\\
        Veg Momos & 4\\
        Veg Fried Rice & 4 \\
        Rajma & 3 \\
        Naan & 4 \\
        Dal Makhani & 5 \\
        Masala Dosa & 5 \\
        Palak Paneer & 3 \\
        Khakhra & 2 \\
        Malai Kofta & 3 \\
    \bottomrule
    \end{tabular}
    \end{table}
    

    
    \begin{table}
        \caption{Results on training data}
        \centering
        \begin{tabular}{cccc}
        \toprule
             Method & RMSE \\
        \midrule
             Matrix Factorisation &  1.030 \\
             TF-IDF & 1.040 \\
             TF-IDF with flavour & 1.041 \\
        \bottomrule
        \end{tabular}
        \label{tab:results-training}
    \end{table}
    
    \begin{table}
        \caption{Results of online A/B testing}
        \centering
        \begin{tabular}{cccc}
        \toprule
             Method & RMSE \\
        \midrule
             Matrix Factorisation & 2.93 \\
             TF-IDF & 2.11 \\
             TF-IDF with flavour & 1.94 \\
        \bottomrule
        \end{tabular}
        \label{tab:results-testing}
    \end{table}
    
    When making recommendations for a particular user, the Collaborative Filtering Algorithm only considers other similar users. It does not take into account the content or features of an item, hence, food flavour cannot be incorporated when using this method to make recommendations.
    \\
    \item \textbf{Content Based Filtering:}
    
    Content Based Filtering algorithm takes into consideration the content or properties of an item. Items are described with a set of descriptor terms or tags whichform the basis for item-based comparison.
    
    Term Frequency Inverse Document Frequency (TF-IDF) has its roots in Information Retrieval but finds its application in Content Based Recommendation Systems. We describe food dishes using their ingredients, the cuisine and whether the dish is vegetarian or non-vegetarian. A few examples of tags associated with some dishes are:
    \begin{itemize}
        \item
            \textquotesingle
                Aloo Paratha
            \textquotesingle - [vegetarian, cumin, flour, ginger, lemon, masala, oil, paratha, potato, salt, wheat, north indian, punjab]
        
        \item
            \textquotesingle
                Palak Paneer
            \textquotesingle
                - [vegetarian, clove, coriander, cumin, curry, garlic, ginger, masala, paneer, salt, spinach, tomato, turmeric, north Indian, Punjab]
    \end{itemize}
    
    We associated the 1381 dishes in our database with 397 unique tags as described above. Next, for each tag associated with a dish, the TF-IDF scores for the tag was calculated using the standard TF-IDF formula.
    
    \begin{equation}
        \label{tfidf_formula}
        W_{x, y}={tf}_{x, y}\cdot\log{(\frac{N}{df_x})}
    \end{equation}
    Where, \\
    \begin{itemize}
        \item x is the set of tags and y is the set of dishes\\
        \item $tf_{x, y}$ = 1 if dish y has tag x else 0 \\
        \item $df_x$ = number of dishes containing tag \\
        \item N = total number of dishes
    \end{itemize}
    
    For each dish, the TF-IDF calculation results in the formation of a vector of length 397. A slice of such a vector is shown in Table \ref{tab:vector_values}.

    For a given user, equation (\ref{preference_tf_idf}) is used to calculate the preference score for an unrated dish \textit{i}, using the scores for all dishes \textit{j} that the user has rated, by calculating the similarity between dish \textit{i} and \textit{j}, and weighing the cosine similarity with the dish score \textit{j}.
    Similarly, as Table \ref{tab:vector_values_with_flavour} depicts, user preference for dishes is also calculated after including 5 dish flavours and performing a weighted average of the cosine similarity between the dish's ingredient descriptors and flavour descriptors.
    Assume $\alpha$ such that $\alpha\in$ dishes rated by the user and $\beta$ to be \textit{CosineSimilarity(i,j)}. Then the user preference score will be:
    \begin{equation}
        \label{preference_tf_idf}
        \forall i \in All\ dishes\ in\ the\ database\frac{\sum_{\alpha}^{}\cdot \beta\cdot UserScore(i)}{\sum_{\alpha}^{}\beta}
    \end{equation}

\end{enumerate}

\section{Results}
We consider three recommendation systems, the first one being an implementation of collaborative filtering (Matrix Factorization) and the second and third being implementations of content based filtering (TF-IDF), with and without flavour incorporated respectively. The users previous ratings as seen in Table \ref{tab:user_reviewed_dishes} revolve around a lot of rice based dishes as well as rich dishes. Matrix factorization predicts dishes based on the patterns of other users. This however fails to consider the health aspect. TF-IDF improves upon this and suggests much healthier alternatives but erroneously predicts what a user may like. TF-IDF with flavour is the best of both worlds where it predicts healthy food items while maintaining an acceptable level of user taste preferences. Quantitatively Table \ref{tab:results-training} shows the Root Mean Squared Error on the training data which fails to decisively showcase which method is the best. However, the RMSE for our online A/B testing between three various recommendation approaches, as shown in Table \ref{tab:results-testing} yields much better results as reasoned in the qualitative analysis and therefore, we conclude that using TF-IDF with flavour improves recommendation. 

We developed a rudimentary cuisine classifier using Naive-Bayes to assign cuisines to dishes based on their ingredients. However, the set of dishes we were working with had a heavy bias towards North Indian dishes, and thus had a insignificant impact on the quality of recommendations.

\section{Future Work}
The quality of recommendations could be significantly improved with the incorporation of a cuisine element. However,
this will require our data set to expand to a multitude of cuisines. The preparation techniques will also need to be
considered during classification, as it varies from cuisine to cuisine. 

Contextual factors such as seasonal sensitivity could be incorporated into the system, that adds an element of personalizing based on more context. For example, people enjoy hot drinks in cold weather, or cold drinks in hot weather. Another example - a novelty function could be incorporated which considers the seasonal trend of a users preferences which can then be used to fine-tune the flavour scores. 

Significant improvements can be made to the flavour profiling, with the availability of complete and accurate nutritional data for store-bought foods. Existing regulations do not mandate all suppliers to report such data in detail. However, other flavour scores could be refined if the data is present. Our database of foods can grow considerably larger as a result of this.

We initially were inspired by encouraging people to eat healthier through better recommendations. The limited scope of this work is on food flavour profiling, and we hope to bolster the work and accelerate the research field by more thorough integration of health factors into food recommendation at large.

Ultimately, the body of food computing research in understanding how item taste needs to be explored in much greater detail. This may be through a combination of more detailed databases including chemical profiles and quantitative metrics to understand flavours. In parallel, understanding the taste preferences at an individual level needs to be captured. Accurate personal food logging is a notable challenge in this area that will help fuel better user understanding. We hope the above preliminary research efforts aid in the momentum towards better quality of life through healthy and enjoyable culinary experiences.

%

%
\bibliographystyle{ACM-Reference-Format}
\bibliography{sample-base}

%

\end{document}